\documentclass{elsart}

 \usepackage{graphicx}

\usepackage{amssymb}

\journal{Physica A}

\begin{document}

\begin{frontmatter}



\title{Monte-Carlo analysis of critical properties of the two-dimensional
randomly site-diluted Ising model via Wang-Landau algorithm}


\author{I. A. Hadjiagapiou\corauthref{cor1}},
\corauth[cor1]{Corresponding author.}
\ead{ihatziag@phys.uoa.gr}
\author{A. Malakis},
\author{S. S. Martinos}

\address{Section of Solid State Physics, Department of Physics,
University of Athens, Panepistimiopolis, GR 15784 Zografos,
Athens, Greece}

\begin{abstract}

The influence of random site dilution on the critical properties
of the two-dimensional Ising model on a square lattice was
explored by Monte Carlo simulations with the Wang-Landau sampling.
The lattice linear size was $L = 20-120$ and the concentration of
diluted sites $q=0.1, 0.2, 0.3$. Its pure version displays a
second-order phase transition with a vanishing specific heat
critical exponent $\alpha$, thus, the Harris criterion is
inconclusive, in that disorder is a relevant or irrelevant
perturbation for the critical behavior of the pure system. The
main effort was focused on the specific heat and magnetic
susceptibility. We have also looked at the probability
distribution of susceptibility, pseudocritical temperatures and
specific heat for assessing self-averaging. The study was carried
out in appropriate restricted but dominant energy subspaces. By
applying the finite-size scaling analysis, the correlation length
exponent $\nu$ was found to be greater than one, whereas the ratio
of the critical exponents ($\alpha /\nu$) is negative and ($\gamma
/\nu$) retains its pure Ising model value supporting weak
universality.

\end{abstract}
\date{\today}

\begin{keyword}
Lattice theory \sep two-dimensional Ising model\sep randomness
\sep site dilution \sep Monte-Carlo \sep Wang-Landau \sep
finite-size scaling

\PACS 05.50.+q
\sep75.10.Hk \sep 05.10.Ln \sep 64.60.Fr
\end{keyword}
\end{frontmatter}

\newpage
\section{Introduction}

In the last few years, many theoretical, numerical and
experimental investigations have appeared for the study of the
influence of disorder (usually quenched site or bond dilution) on
statistical systems in the presence or absence of an external
magnetic field, since an experimental sample is not free of
disorder, in general. Such systems are randomly dilute uniaxial
antiferromagnets, e.g., $Fe_{q}Zn_{1-q}F_{2}$,
$Mn_{q}Zn_{1-q}F_{2}$, $Fe_{1-q}Al_{q}$, obtained by mixing
uniaxial antiferromagnets with non-magnetic materials. These
systems are modelled by using pure systems models modified
accordingly. The most popular models for pure systems are those of
Ising and Potts; in the current case, the former one shall be
used.

The Ising model, a favorite for physicists, is used as the
prototype for phase transitions, critical phenomena, biological
and econophysics, owing to the fact that its two-dimensional
version on a square lattice was solved analytically by Onsager,
\cite{onsager}; it has also become a standard model for testing
scaling and universality hypotheses. The Ising model as well as
that of Potts, Heisenberg, Baxter-Wu, etc., represent collective
phenomena which are difficult to be solved exactly, except in some
cases, consequently approximate methods have been developed to
cope with, such as mean field theory, numerical methods,
perturbation theory, scaling, renormalization group, Monte Carlo
etc. The effect of randomness on the critical behavior and
magnetic phase diagrams of classical random spin systems has
attracted much interest in recent years. Randomness is encountered
in the form of vacancies, variable bonds, impurities, random
fields, etc. Its influence on the critical behavior and magnetic
properties is a long-standing and still unsettled problem in
statistical physics. Similar methods, as in the pure systems, have
been utilized for the study of these systems. In these studies, an
important question which had arisen was the extent to which
randomness influences the critical behavior and magnetic
properties. The first remarkable criterion for the influence of
randomness on a critical system was proposed by Harris,
\cite{harris}, according to which randomness changes the critical
behavior if the specific heat critical exponent $\alpha$ of the
pure system is positive, $\alpha> 0$. In this case a new critical
point with conventional power law scaling and new exponents
emerges. However, the pure two-dimensional Ising model ($2$D IM)
constitutes a marginal situation since $\alpha = 0$. This
exception makes $2$D IM of particular interest attracting much
attention. However, this effort led to contradicting results,
increasing confusion and revealing the inhibit hidden complexity,
implying the need for more subtle approaches to tackle it.

The $2$D IM consists of an array of $N$ fixed points lying on the
sites of a two-dimensional lattice of linear size $L$ such that $N
= L^{2}$. Each lattice site is occupied by a magnetic atom
characterized by the spin variable $ S_{i}, i=1,2,\ldots,N$, with
$S_{i}$ taking on the values $\pm1$. One can also consider a
modified version of this model wherein some of the lattice sites,
chosen randomly, are either vacant or occupied by non-magnetic
atoms, (e.g., Al atoms \cite{alum}); both cases of randomness are
treated equivalently. This version of Ising model is called
two-dimensional randomly site-diluted Ising model ($2$D RSDIM).
For this system, the critical exponents associated with the random
fixed point have been estimated for the dilution-type disorder by
theoretical and numerical approaches, leading to questionable
conclusions, \cite{shchur}. However, these approaches have shed
light on estimates of the critical temperature and nature of the
phase transition. In addition, the phase diagram is strongly
influenced, among other factors, by dilution,
\cite{restrepolandau}. The concentration of vacancies (or
nonmagnetic particles) is denoted by $q$ (dilution), while that of
occupied sites (magnetic particles) by $p$ (purity), $q + p =1$.
The vacancies are considered to be quenched and uncorrelated,
since one can also encounter systems where the vacancy locations
are correlated, \cite{thurston,weinrib}.

The process to be followed here for tackling the RSDIM is that of
Monte Carlo. The MC approach has been proved to be a powerful tool
to study difficult problems such as random spin systems. In some
cases, the simulation method suffers from problems of slow
dynamics, thus, new algorithms have been proposed to overcome such
difficulties. Wang--Landau (WL)\cite{wanglandau} and entropic
sampling \cite{lee} are examples of such efforts. The critical
properties concern always an infinitely bulk system, since the
phase transitions appear in such a system; however, from MC
simulations the critical behavior is extracted from results
obtained on a finite-size system by means of finite-size scaling
(FSS). This process, since its inception, has been evolved as a
very powerful tool for extracting properties of an infinite system
near a phase transition, although it is, as yet, not fully
completed causing, sometimes, ambiguities about its results. The
major goal of the finite-size method is to identify the set of
critical exponents that, together with other universal parameters,
characterizes the universality class. As these exponents offer the
most direct test of universality, their precise calculation is of
great importance. However, the experimental devices are finite,
consequently, the exponents cannot be measured with infinite
precision causing, occasionally, controversies to distinguish the
universality class a specific system belongs, \cite{fisher1}. For
more on the theory of finite-size scaling see, e.g., Barber
\cite{barber}, Privman \cite{privman} and Binder \cite{binder1}.

For the $2$D RSDIM, the nature of the possible phase transition as
well as the universality class is not completely understood. The
value and even the sign of the specific heat exponent $\alpha$ is
still not known. Because of logarithmic divergence of specific
heat of the pure system ($\alpha = 0$), Harris criterion is
inconclusive, hence, a great deal of effort has been dedicated to
elucidating the properties of the $2$D RSDIM. Currently, it seems
that two scenarios have prevailed, which, however, are mutually
exclusive. According to the first one
(\textit{logarithmic-corrections} scenario), the critical
exponents are unaffected by disorder, apart from possible
logarithmic corrections (\textit{strong-universality}), while the
second, predicts critical exponents varying continuously with
disorder, but the exponents' ratios $ (\gamma/\nu) $, $ (\beta/
\nu) $ remain the same as in the pure case, \textit{weak
universality}, \cite{shchur}.

Kim and Patrascioiu \cite{kimpatra} studied the $2$D RSDIM on a
square lattice with periodic boundary conditions and dilution-site
concentration (dilution probability) $q = 1/9, 1/4, 1/3$ and
lattice linear size $L$ up to $600$, using MC simulation; their
conclusions are that the specific heat does not diverge for $q =
1/4, 1/3$ while for $q = 1/9$ it seems to increase as $\epsilon
\rightarrow 0$, $\epsilon = (T-T_{c})/T_{c}, T_{c}$ critical
temperature. The magnetic susceptibility $\chi$ and correlation
length $\xi$ fit the pure power law, the value of the respective
exponent $\gamma$ and $\nu$ increases with $q$ while $\eta = 2-
\gamma/\nu$ remains the same as in the pure system. Queiroz and
Stinchcombe \cite{questinch} using transfer-matrix-scaling
technique, Mazzeo and K\"{u}hn \cite{mazzeokuhn} following the
same technique with the equilibrium ensemble approach to
disordered systems, came to the same conclusions. Newman and
Riedel \cite{newmanriedel}, using renormalization group on weakly
diluted systems confirmed the existence of a new stable fixed
point with new exponents. Heuer \cite{heuer2d} focused on the
exponents $(\gamma / \nu)$ and $(\beta / \nu)$ for the $2$D RSDIM
on a square lattice with $72 \leq L \leq 250$ and $0 \leq q \leq
0.4$; according to his estimations, $(\beta / \nu)$ does not
change notably with dilution within the errors and, practically,
it assumes its pure system value. The exponent $(\gamma / \nu)$
shows a strong dilution-dependence in the temperature range $
10^{-2} < (T-T_{c})/T_{c} < 1 $, but near $T_{c}$ it
asymptotically approaches the pure system value $(7/4)$
independently of dilution.

On the other hand, Shchur and Vasilyev \cite{shchur} analyzing the
MC data for the magnetic susceptibility critical amplitudes
$(\Gamma, \Gamma')$ above and below the critical temperature,
respectively, for dilution $q \leq 0.25$ and lattice linear size
up to $L = 256$, concluded that the ratio $(\Gamma / \Gamma')$
seems to remain constant for the dilute-site concentrations
considered and equal to its pure system value. This implies that
the $2$D RSDIM belongs to the same universality class as the pure
one. Dotsenko and Dotsenko \cite{dotsenkodotsenko}, Shalaev
\cite{shalaev}, Shankar \cite{shankar}, Ludwig \cite{ludwig},
using field theoretical calculations, showed that randomness is
irrelevant (critical exponents are unchanged) and only logarithmic
corrections might appear in the case of weak dilution; they found
for the correlation length, magnetization and susceptibility,
respectively,

\begin{equation}
   \xi \propto |\epsilon| ^{-1} [1+ \lambda \ln(1/|\epsilon|)]^{1/2}
         \;\; \label{xi}
\end{equation}

\begin{equation}
   m \propto |\epsilon|^{1/8}[1+ \lambda \ln(1/|\epsilon|)]^{-1/16}
      \;\; \label{magn}
\end{equation}

\begin{equation}
   \chi \propto |\epsilon| ^{-7/4} [1+ \lambda \ln(1/|\epsilon|)]^{7/8}
         \;\; \label{xi}
\end{equation}

\vspace{-5mm}

while the specific heat diverges as ln(lnL)

\begin{equation}
 C_{V} \propto |\epsilon|^{-\alpha} \ln[1+ C
 \ln(1/|\epsilon|)] + C'   \;\; \label{spech}
\end{equation}

\vspace{-5mm}

where $\lambda$ is a smooth function of q, $\alpha = 0$ and $C'$ a
constant. Ballesteros et al \cite{ballesterosetal2}, by performing
MC simulations in conjunction with finite-size scaling (FSS) for
$p=1, 8/9, 3/4, 2/3$, although observed small deviations of the
values of the critical exponents from those of the pure system on
varying the concentration, they considered it as a transient
effect, since this can be lifted if a pure Ising value for the
exponents is combined with logarithmic corrections. Selke et al
\cite{selkeetal}, using MC techniques for lattices with linear
size $8 \leq L \leq256$ and spin concentration $0.1 \leq p \leq
1$, concluded that impurities lead the specific heat to diverge as
$C \sim$ ln(lnL)  on approaching the critical temperature. Tomita
and Okabe \cite{tomitaokabe} performed MC simulations on the same
system on a square lattice using the probability-changing cluster
(PCC) algorithm confirming that randomness is irrelevant and its
influence is evident through logarithmic corrections.

Allowing for the previous arguments, we remark that in spite of
much effort devoted to the investigation of RSDIM in the critical
region, the question of the dependence of critical exponents on
randomness is still open. In this paper, we shall examine the
critical properties of the $2$D RSDIM using an alternative
approach based on the WL algorithm. In this analysis, the major
goal will be to estimate the critical temperature and exponents of
the $2$D RSDIM for various values of the spin concentrations to
assess whether it belongs to the Ising universality class or not
by studying the finite-size behavior of the specific heat and
magnetic susceptibility.

The paper is organized as follows. In the next section, after the
introduction of the model, we shall discuss the approach based on
the WL algorithm and an efficient implementation by conveniently
restricting the simulation in the dominant energy subspace. In
section $3$ we apply the FSS to the Ising model under
consideration and we close with the conclusions and discussions in
section $4$.

\vspace{-7mm}

\section{Numerical approach of the RSDIM}
\noindent

\vspace{-7mm}

We consider the Hamiltonian of the two-dimensional site-diluted
Ising model, in the absence of any external field,

\begin{equation}
 H=-J\sum_{<ij>}c_{i}c_{j}S_{i}S_{j} ,\;\;S_{i}=\pm1, \label{first}
\end{equation}

\vspace{-5mm}

where $J > 0$ is the interaction constant, ferromagnetic
interactions. The coefficients $c_{i}\,'s$, called occupation
variables, are quenched, uncorrelated random variables chosen to
be equal to $1$ with probability $p$, when the $i$-site is
occupied by a magnetic atom and $0$ with probability $q = 1-p$
otherwise; that is, we have the probability distribution
$P(c_{i})=p\: \delta (c_{i}-1) + q\, \delta(c_{i})$. The summation
extends over all nearest-neighbor pairs of the square lattice, of
linear size $L$ with periodic boundary conditions.

The data was generated by extensive MC calculations using WL
sampling method to estimate the density of states (DOS) $g(E)$
\cite{wanglandau}. WL sampling performs a random walk with an
acceptance ratio $P(E_{i}\rightarrow E_{j})=\min{\{1, [g(E_{i}) /
g(E_{j})]\}}$, $E_{i}$ and $E_{j}$ are the energies before and
after the transition, respectively, aiming at sampling a flat
histogram in energy. The WL algorithm overcomes the difficulties,
such as critical slowing down and long relaxation times in systems
with complex energy landscape, appearing in other MC processes.
This algorithm has been applied to a wide range of systems
spanning from the Ising model \cite{12,13}, random field Ising
model \cite{malakisfytas}, $3$D conserved-order-parameter Ising
model \cite{hadjiagapiou}, Potts \cite{14}, to glassy systems
\cite{15}, polymers \cite{16}, DNA \cite{17} and the Baxter-Wu
model \cite{18}. The DOS $g(E)$ is not constant during the random
walk, it is modified according to the rule $g(E) \rightarrow
(g(E)\cdot f)$; the modification factor $f$ varies as
$f_{j+1}=\sqrt{f_{j}}$, $j$ is the order of iteration. In the
current investigation, the WL algorithm performed $26$ iterations
and the initial value for $f$ was $f = e$.

Having an accurate estimation of $g(E)$, the non-normalized
canonical distribution can be constructed, $P(E,T) = g(E)
e^{-\beta E}$; subsequently, the partition function can be
calculated through the expression $Z(T) = \sum_{E} g(E) e^{-\beta
E}$, from which most of the thermodynamic observables can be
estimated. This kind of MC simulation constitutes a major
improvement towards speeding calculations since it avoids multiple
runs (one for each temperature) needed by the majority of MC
algorithms to describe the temperature dependence of thermodynamic
quantities over a significant temperature range; in a WL algorithm
temperature is not needed to be specified a priori. In the present
case, the WL algorithm was implemented on lattices with $20 \leq L
\leq 120$ and the density of states was stored as a function of
the energy. The dilution $q$ can vary from $0.0$ $(p = 1)$ to the
percolation threshold $q _{c}^{PERC} = 0.407255$ $(p_{c}^{PERC} =
0.592745(2))$, \cite{ziff}.

The quantities, upon which we shall rely for studying the critical
behavior of the RSDIM, are the specific heat
$\left(C=([<E^{2}>]-[<E>]^{2})/(NT^{2})\right)$ and magnetic
susceptibility $\left(\chi=([<M^{2}>]-[<M>]^{2})/(NT)\right)$, per
particle. The MC data is generated by choosing a realization of
the dilution for a specific value of $q$ and various values of
$L$. This procedure is then repeated for several other
realizations for a specific value for the lattice linear size $L$.
For each realization, we find the maximum value of the specific
heat $(C^{*}(q,L))$ and susceptibility $(\chi ^{*}(q,L)$),
recording simultaneously the values of the respective
pseudocritical temperatures, $T^{*}_{C}(q,L)$ and $T
^{*}_{\chi}(q,L)$, respectively, forming in each case, a sequence
of ``pseudocritical temperatures" converging to the critical
temperature $T_{c}(q)$ of the infinite system as $L \rightarrow
\infty$. Among the different realizations large fluctuations are
observed in the same set of the previous quantities.

The presence of randomness is evident in the way averaging
processes are carried out for an observable $X$, which assumes a
different value for each of the $M$ random realizations of the
disorder corresponding to the same value of dilution q. This
implies that $X$ behaves as a stochastic variable, whose mean
value is estimated through a two-step process. First, the usual
thermal average is performed for a specific realization of the
dilution. After the completion of the M realizations of
randomness, the disorder average is carried out over the M
realizations and is denoted by the brackets [ ].

For the $2D$ IM the specific heat and magnetic susceptibility in
the thermodynamic limit diverge; in a finite lattice, this
divergence is rounded off and manifests itself by a maximum
exhibited by the above quantities. This maximum increases
gradually with $L$ and ultimately tends to infinity as $ L
\rightarrow\infty $. For the $2$D RSDIM, in attempting to detect
the maximum $C^{*}(q,L)$ of the specific heat and the respective
temperature $T^{*}(q,L)$ (pseudocritical temperature), we
considered two routes. Let $C_{m}(q,L)$ be the specific heat for a
particular realization $m$ out of $M$ realizations for a specific
dilution $q$. In the first route, we estimated, initially, the
maximum value $C^{*}_{m}(q,L)$ together with the respective
pseudocritical temperature $T^{*}_{C,m}(q,L)$ for every
realization $m$ of the disorder; then, we considered the
\textit{sample average} of the individual specific heat maximum
$[C^{*}(q,L)]$ and pseudocritical temperature $[T^{*}_{C}(q,L)]$,
for the $M$ realizations,

\begin{equation}
 [C^{*}(q,L)] = \frac{1}{M}
 \sum_{m=1}^{M}C^{*}_{m}(q,L),\;\;\;\;\;\;
  [T^{*}_{C}(q,L)] = \frac{1}{M} \sum_{m=1}^{M}T^{*}_{C,m}(q,L) \;\; \label{specheat1}
\end{equation}

In the second route, following Rieger and Young, the
\textit{sample summation} for the specific heat for the totality
of $M$ realizations was considered \cite{riegeryoung},

\begin{equation}
 [C(q,L)]_{sum} = \frac{1}{M}
 \sum_{m=1}^{M}C_{m}(q,L)  \;\;\;\; \;\; \label{specheat2}
\end{equation}

In (\ref{specheat2}), the resulting specific heat curve is very
complex with many local maxima, reflecting the strong
pseudocritical temperature fluctuations in the ensemble of random
realizations. From these maxima we selected the absolute one,
indicated by $[C(q,L)]^{*}_{sum} \equiv max[C(q,L)]_{sum}$,
occurring at the pseudocritical temperature $T^{*}_{C,sum}(q,L)$.
The same procedure was also followed for the magnetic
susceptibility $\chi$; the resulting quantities are denoted by
$[\chi^{*}(q,L)]$, $[T^{*}_{\chi}(q,L)]$, $[\chi(q,L)]^{*}_{sum}$,
$T^{*}_{\chi,sum}(q,L)$.

To investigate the critical behavior of the system, we performed
extensive MC simulations to calculate the density of states $g(E)$
for each value of dilution $q$ and $L$ through the WL algorithm.
After estimating the density of states $g(E)$, one can proceed to
the calculation of the necessary thermodynamic quantities, such as
the energy $E$, specific heat $C_{L}(T)$, magnetization $M$ and
susceptibility $\chi_{L}(T)$ for further use in order to identify
the probable universality class. However, before proceeding to
this end, we outline a new restricted method for speeding up the
numerical calculations; this method is called ``\textit{Critical
Minimum Energy Subspace}" (CrMES) technique,
\cite{13,hadjiagapiou,18}. In this method, we concentrate our
simulation only on the dominant energy subspaces. In the following
lines we describe its implementation. Consider the specific heat
per site for a lattice of linear size $L$ at temperature $T$,

\vspace{-5mm}

\begin{eqnarray}
 C_{L}(T)=L^{-d}T^{-2} \mbox{\Huge \{ }
 Z^{-1} \sum_{E_{min}}^{E_{max}} E^{2} \exp[S(E)-\beta E]-   \nonumber  \\
  \mbox{\LARGE(}
  Z^{-1} \sum_{E_{min}}^{E_{max}} E \exp[S(E)-\beta E]
 \mbox{\LARGE )}^{2}
 \mbox{\Huge \} }      \label{second}
\end{eqnarray}

\vspace{-5mm}

where the Boltzmann constant was set $k_{B} =1$, thus $\beta =
1/T$, $d$ is the spatial dimension $(d = 2)$ and Z the ``partition
function" of the system,

\begin{equation}
  Z  =  \sum_{E_{min}}^{E_{max}}\exp[S(E)-\beta E] \;\;\;\; \label{third}
\end{equation}

\vspace{-5mm}

The latter expression is the partition function in case $g(E)$ is
the exact DOS of the system and properly normalized, \cite{AJP72}.
In practice, the DOS, resulting from WL simulations, is an
approximate result whose accuracy depends on that of the
simulation. In the expression (\ref{second}), the calculation of
the specific heat in the critical region can be speeded up by
restricting the energy interval if we use the CrMES technique. Let
$\tilde{E}$ be the energy corresponding to the maximum term
$exp[S(E)-\beta E]$ of the partition function (\ref{third}) for
the temperature at hand. Because of the sharpness of the energy
distribution, the energy interval $(E_{min},E_{max})$ in the
summation (\ref{second}) is replaced by a smaller one
$(\tilde{E}_{-},\tilde{E}_{+})$ around $\tilde{E}$ corresponding
to a predefined accuracy $r$ for the specific heat expressed as,
$\mid[C_{L}(\widetilde{E}_{-},\widetilde{E}_{+})/C_{L}(E_{min},E_{max})]-1\mid\leq
r$, where $r=1\cdot10^{-6}$ and
$\widetilde{E}_{\pm}=\widetilde{E}\pm\Delta _{\pm}$, $\Delta
_{\pm} \geq 0$. The induced errors are much smaller than the ones
in determining the DOS, for more see \cite{13}. The magnetic
properties were obtained using the final stages of the WL
algorithm, following our earlier proposal as in \cite{mmhfk}.

\vspace{-5mm}

\section{Finite-size scaling analysis. Results}

\noindent

\vspace{-5mm}

The FSS is based on the assumption that the free energy of a
system of linear size $L$ and in the absence of an external
magnetic field scales as,

\begin{equation}
  F(L,\epsilon)  =  L^{-\psi} F_{0}(\epsilon L^{\theta}) \;\;\;\; \label{forth}
\end{equation}

where $\psi =(2- \alpha)/\nu$. The scaling of the correlation
length $\xi = \xi_{0} \epsilon^{-\nu}$ suggests that $\theta =
\nu^{-1}$. The scaling function $F_{0}(x)$ is universal, in that,
it is independent of the lattice size.

\begin{figure}[htbp]
\includegraphics*[height=0.4\textheight]{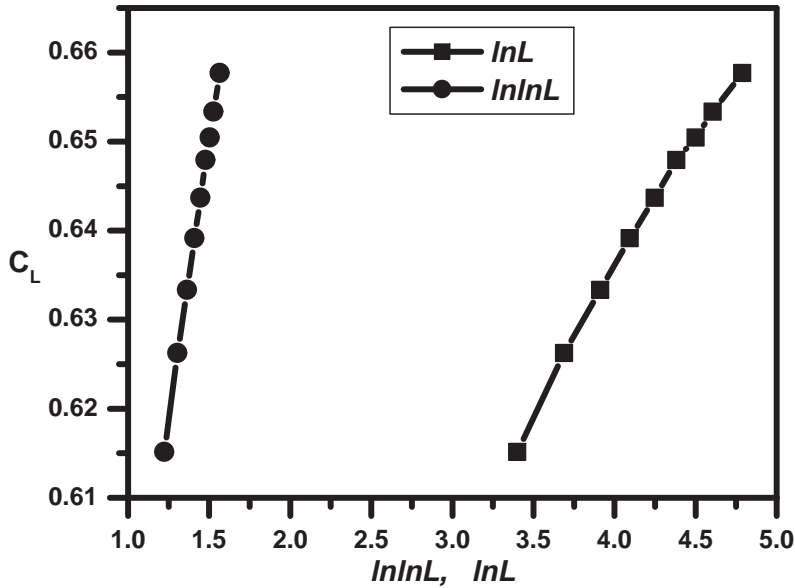}
 \caption{\label{Fig1} Specific heat finite size approximations
 against $lnL$ (squares) and $lnlnL$ (circles) for $q = 0.3$.}
\end{figure}

The quantity that specifies the scenario the RSDIM shall follow
is, nevertheless, the specific heat. According to the lnlnL
scenario the specific heat diverges as in (\ref{spech}), while
according to weak universality and FSS, the specific heat obeys
the power law,

\begin{eqnarray}
[C^{*}(q,L)] = p_{1}+ q_{1}L^{\alpha / \nu_{C}} \;\;\;\;
\label{specsamplav}
\end{eqnarray}

In this expression the lack of the correction terms speeds up the
convergence and yields more stable fits.

The choice of the suitable asymptotic law for the specific heat
has caused considerable concern, since one has to choose between
Eqs. (\ref{spech}) and (\ref{specsamplav}); thus, we have checked
both. In the pure two-dimensional Ising model, $q = 0$, the
specific heat against lnL is a straight line, representing the
forthcoming divergence, while against lnlnL the respective curve
bends upwards, see Fig. $6$ in \cite{mazzeokuhn} as well as Ref.
\cite{staufferetal}. Plotting the specific heat data for $q=0.3$
against lnL and lnlnL, respectively, see Fig.~\ref{Fig1}, we
observed that the former curve (rightmost in Fig.~\ref{Fig1})
bends downwards deviating from a straight line, whereas the latter
curve (leftmost) seems to bend also downwards implying that the
data does not follow the lnlnL scenario. The same behavior was
also observed for the dilution $q=0.2$, but this is more evident
for the stronger dilution $q=0.3$; for $q=0.1$ it is not so
evident because this is a crossover case.

Mazzeo and K\"{u}hn (\cite{mazzeokuhn}) checking the credibility
of both scenarios, studied initially the lnlnL one; by fitting
their data to this law, they concluded that it was difficult to
accept definitely this scenario (\ref{spech}); instead, they
focused on the power law (\ref{specsamplav}) and estimated the
$\alpha$-exponent for various dilutions, testing the validity of
the respective values by using the hyperscaling relation in
combination with the $\nu$-exponent, estimated earlier. Their data
resulted from calculations on the equilibrium ensemble approach
together with numerical transfer matrix technique,
phenomenological renormalization group scheme and conformal
invariance on finite-width strips.

In addition, we have also invoked the $\chi ^{2}$-test to
discriminate between Eqs. (\ref{spech}) and (\ref{specsamplav}) by
estimating the ratio between the $\chi ^{2}$-test for the lnlnL
scenario with the one for the power law suggested by
(\ref{specsamplav}): for $q=0.2$ this ratio is $2$ while for
$q=0.3$ it is $3$, indicating an increasing tendency with $q$;
hence, although the above description makes evident the need for
data on much larger systems sizes, we shall adopt the power law
(\ref{specsamplav}) as it seems to be more suitable and reliable
for the present case. This practice shall be followed in the
sequel.

\begin{table}
\caption{\label{tablesample}The critical temperature and critical
exponents resulting from the sample average of the specific heat
(\ref{specheat1}) and the respective expression for
susceptibility, for different values of spin dilution $q$. For
each q-value, the first line corresponds to the specific heat data
and the second to the susceptibility.}
\begin{tabular}{ccccccc}
\hline $q$ &  $T_{c}$ & $\nu$  & $\alpha / \nu$  & $C^{\infty}$ &
$\gamma / \nu$   \\  \hline
$0.1$  & $1.90668(0.00101)$ & $1.15209^{0.03699}_{0.03476}$ & $-0.26406(0.02552)$ & $2.38683(0.09096)$    \\
       & $1.90271(0.00219)$ & $1.15181^{0.03813}_{0.03576}$ &            &        & $1.74968(0.01369)$  \\
$0.2$  & $1.50675(0.00137)$ & $1.18209^{0.02647}_{0.02534}$ & $-0.30892(0.01719)$ & $1.24453(0.01400)$   \\
       & $1.50180(0.00342)$ & $1.18289^{0.04776}_{0.04419}$ &            &        & $1.74918(0.00983)$  \\
$0.3$  & $1.15280(0.00385)$ & $1.23629^{0.05739}_{0.05251}$ & $-0.38223(0.03119)$ & $0.71851(0.00659)$    \\
       & $1.10051(0.00211)$ & $1.23718^{0.02949}_{0.02815}$ &            &        & $1.74939(0.01643)$  \\
\hline \hline
\end{tabular}
\end{table}

\begin{table}
\caption{\label{tablesummation}The critical temperature and
critical exponents resulting from the sample summation for the
specific heat (\ref{specheat2}) and the respective expression for
susceptibility, for different values of spin dilution $q$. For
each q-value, the first line corresponds to the specific heat data
and the second to the susceptibility.}
\begin{tabular}{ccccccc}
\hline $q$ &  $T_{c}$ & $\nu$  & $\alpha / \nu$  & $C^{\infty}$ &
$\gamma / \nu$   \\  \hline
$0.1$  & $1.89557(0.00218)$ & $1.15335^{0.06633}_{0.05948}$ & $-0.26412(0.02986)$ & $2.29955(0.0988)$    \\
       & $1.90329(0.00322)$ & $1.15303^{0.05880}_{0.05336}$ &       &             & $1.74934(0.00694)$  \\
$0.2$  & $1.49638(0.00272)$ & $1.18318^{0.04952}_{0.04569}$ & $-0.30877(0.01983)$ & $1.18223(0.01361)$   \\
       & $1.50281(0.00387)$ & $1.18293^{0.05486}_{0.05021}$ &       &             & $1.74931(0.00963)$  \\
$0.3$  & $1.08592(0.00621)$ & $1.23752^{0.06445}_{0.05837}$ & $-0.38221(0.03831)$ & $0.53077(0.00704)$    \\
       & $1.07301(0.00426)$ & $1.23634^{0.05141}_{0.04746}$ &       &             & $1.74925(0.02756)$  \\
\hline \hline
\end{tabular}
\end{table}


\begin{figure}[htbp]
\includegraphics{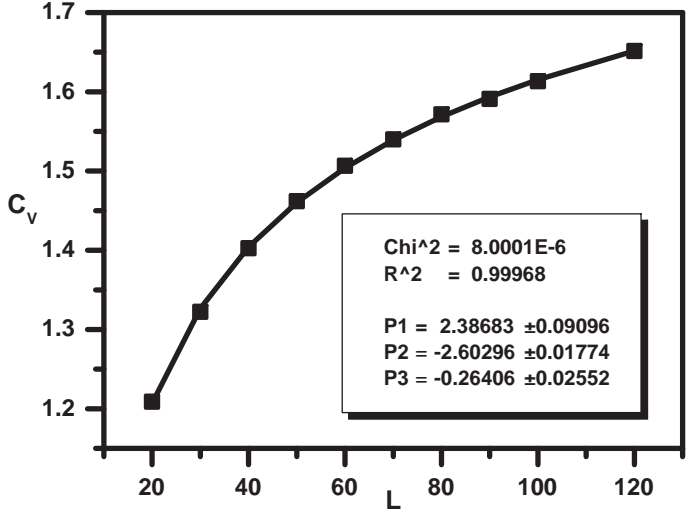}
\caption{\label{Fig2} Specific heat fitting according to the FSS
prediction (\ref{specsamplav}). The specific heat corresponds to
the sample average (\ref{specheat1}) and site dilution $q=0.1$.
For $q=0.2, 0.3$ the plots are similar. The coefficients
$P_{1},P_{2},P_{3}$ correspond to $p_{1}, q_{1}, \alpha /
\nu_{C}$, respectively, see Eq. (\ref{specsamplav}).}
\end{figure}


Firstly, we fit the specific heat sample average maxima to the
power law (\ref{specsamplav}). Considering a specific value for
$q$ and extrapolating towards $L\rightarrow\infty$ the respective
values of $[C^{*}(q,L)]$, one can read off the asymptotic value
$p_{1}\equiv C^{\infty}$ as well as the critical exponents ratio
$(\alpha / \nu)$; these appear in Table ~\ref{tablesample} for the
dilutions $q=0.1, 0.2, 0.3$. We have also considered that the
maximum values $[C(q,L)]^{*}_{sum}$ of the specific heat resulting
form sample summation, see (\ref{specheat2}), follow a similar
power law as in (\ref{specsamplav}), but with different symbols
for the respective quantities,

\vspace{-5mm}

\begin{eqnarray}
[C(q,L)]^{*}_{sum} =   \widetilde{p}_{1}+
\widetilde{q}_{1}L^{\widetilde{\alpha} / \widetilde{\nu}_{C}}
\;\;\;\; \label{specsummav}
\end{eqnarray}

\vspace{-5mm}

Fitting the respective values to (\ref{specsummav}) and
extrapolating towards $L\longrightarrow\infty$, the values for
$\widetilde{p}_{1}$ and $(\widetilde{\alpha} /
\widetilde{\nu}_{C})$ are estimated and appear in Table
~\ref{tablesummation} for the same $q$ values. In both Tables, the
asymptotic values of the specific heat form two decreasing
sequences as $q$ increases and it seems that they tend to zero at
the percolation threshold $(q_{c}=0.407255)$; the non-divergence
is also evident by the levelling off of the specific heat data,
indicative of approaching a saturation value, see Fig ~\ref{Fig2};
the specific heat follows a similar behavior for the other two
values of $q$, as well. The main outcome of this fit is that the
exponents' ratio ($\alpha / \nu$) appears to be negative,
exhibiting a steady decreasing tendency approaching the
percolation value $\alpha / \nu = -0.5$, \cite{nijs}.

For any value of $q$, the respective critical temperature $T_{c}$
is not known a priori (as it happens to be for the random bond
counterpart \cite{aarao}) although there exists a formula
$T_{c}(q) = T_{c}(0) (1 - 1.565q)$, $T_{c}(0)$ is the pure system
critical temperature \cite{stinchcombe}, but of limited
applicability since it is valid only for small impurity
concentrations, thus the estimation of the $T_{c}$ for any $q$ is
of great importance. The correlation length exponent $\nu$ is
directly related to the specific heat exponent $\alpha$ through
the hyperscaling relation $\alpha + d \nu = 2$ ($d$ system's
dimensionality, $d=2$) which acts as a constraint; the
nondivergence of the specific heat ($\alpha < 0$) is consistent
with $\nu > 1$ for the current model. Assuming the FSS prediction,

\begin{eqnarray}
[T^{*}_{C}(q,L)] = T_{c,C} + b_{1}L^{-1/\nu_{C}}  \nonumber  \\
\;\;\;
 [T^{*}_{\chi}(q,L)] = T_{c,\chi} + c_{1}L^{-1/\nu_{\chi}}    \label{TcCCHI}
\end{eqnarray}

\begin{figure}[htbp]
\includegraphics{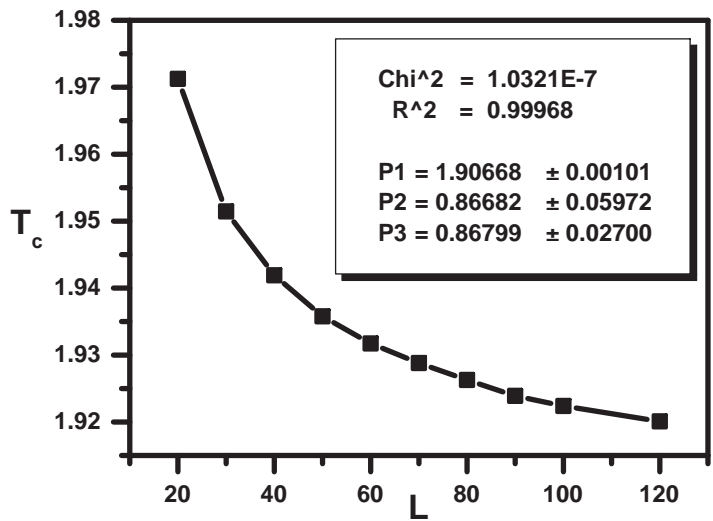}
 \caption{\label{Fig3} Specific heat pseudocritical
temperature fitting according to the first equation of the FSS
prediction (\ref{TcCCHI}). The sequence corresponds to the sample
average (\ref{specheat1}) and site dilution $q=0.1$. For $q=0.2,
0.3$ the plots are similar. The coefficients $P_{1},P_{2},P_{3}$
correspond to $T_{c,C}, b_{1}, (1/\nu_{C})$, respectively, see Eq.
(\ref{TcCCHI}).}
\end{figure}

\vskip 0.5cm

\begin{figure}[htbp]
\includegraphics{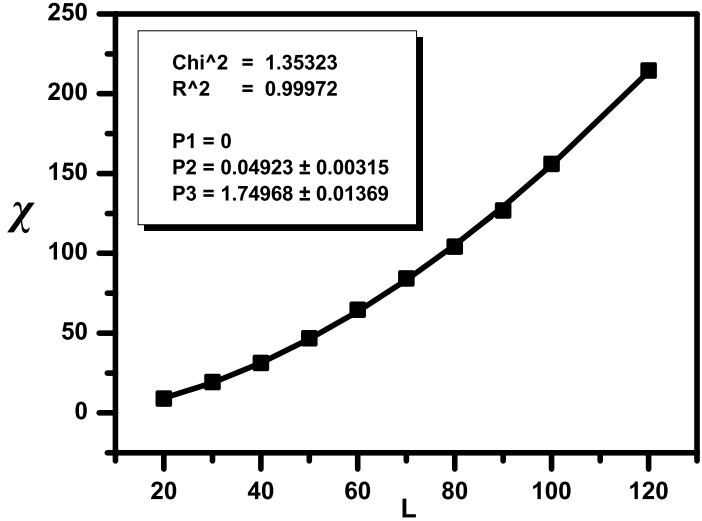}
 \caption{\label{Fig4} Susceptibility fitting according to the
 first equation of the FSS prediction (\ref{chiboth}). The specific
 heat corresponds to the sample average and site dilution $q=0.1$.
For $q=0.2, 0.3$ the plots are similar. The coefficients
$P_{1},P_{2},P_{3}$ correspond to $r_{0}, s_{1}, (\gamma /
\nu_{\chi})$, respectively, see Eq. (\ref{chiboth}).}
\end{figure}

one can estimate the critical temperatures $T_{c,C}$ and
$T_{c,\chi}$ resulting from the sample averages of the specific
heat and susceptibility, respectively, as well as the
corresponding exponent $\nu$, see Fig.~\ref{Fig3}. Also, similar
scaling laws are considered for respective quantities for the
sample summation but with different symbols for the involved
quantities, see \cite{riegeryoung},

\begin{eqnarray}
T^{*}_{C,sum}(q,L) = \widetilde{T}_{c,C} +
\widetilde{b}_{1}L^{-1/\widetilde{\nu}_{C}}  \nonumber  \\
\;\;\;
 T^{*}_{\chi,sum}(q,L) = \widetilde{T}_{c,\chi} +
 \widetilde{c}_{1}L^{-1/\widetilde{\nu}_{\chi}}    \label{TcCCHIsum}
\end{eqnarray}

\vspace{-5mm}

Fitting the respective pseudocritical temperatures for both routes
of the specific heat and susceptibility to (\ref{TcCCHI}) and
(\ref{TcCCHIsum}), they yield the values of the critical
temperature and $\nu$ exponent appearing in
Tables~\ref{tablesample},\hspace{-2mm} ~\ref{tablesummation}. In
both Tables, the respective values for $\nu$ vary continuously
with dilution $q$ and are greater than one in conformity with
hyperscaling. It seems that the critical temperature decreases to
zero as the dilution $q$ increases towards the percolation limit
$q_{c} = 0.407255$. For the low impurity concentration, $q=0.1$,
the deviations of the respective values of the critical
temperature is small, as well as for the intermediate $q=0.2$, see
Tables~\ref{tablesample},\hspace{-2mm} ~\ref{tablesummation}. For
$q=0.1$, the average of the four values is
$\left<T_{c}(q=0.1)\right> = 1.9020625$, which agrees with that in
Heuer \cite{heuer2d} ($1.9004427$), Tomita and Okabe
\cite{tomitaokabe} ($1.9022$), Shchur and Vasilyev \cite{shchur}
($1.9032$), and consistent with that from the formula in Ref.
\cite{stinchcombe}, $T_{c}(q=0.1) = 1.9141$. For the intermediate
impurity concentration $q=0.2$, the average of the respective
critical temperatures is $\left<T_{c}(q=0.2)\right> = 1.5016825$;
it is in agreement with that in Heuer \cite{heuer2d} ($1.507873$),
Shchur and Vasilyev \cite{shchur} ($1.5103$), while deviates from
that of the formula in Ref. \cite{stinchcombe}, $T_{c}(q=0.2) =
1.558930$. For the large impurity concentration $q=0.3$, the first
value of critical temperature in Table~\ref{tablesample}
($1.15280$) shows significant deviation from the other three, so
if we consider only the remaining three their mean value is
$\left<T_{c}(q=0.3)\right> = 1.08648$, consistent with that in
Heuer \cite{heuer2d} ($1.075140$), and Tomita and Okabe
\cite{tomitaokabe} ($1.0712$), while disagrees with that in Ref.
\cite{stinchcombe} $T_{c}(q=0.3) = 1.10306$.

We have also considered the FSS for both susceptibility averages.
The maximum values $[\chi^{*}(q,L)]$ and $[\chi(q,L)]^{*}_{sum}$
for both routes follow the scaling laws,

\vspace{-5mm}

\begin{eqnarray}
  [\chi^{*}(q,L)] = r_{0}+
     s_{1}L^{\gamma / \nu _{\chi}}     \nonumber \\  \;\;\;
[\chi(q,L)]^{*}_{sum} = \widetilde{r}_{0}+
  \widetilde{s}_{1}L^{\widetilde{\gamma} / \widetilde{\nu} _{\chi}}
     \;\;\;\; \label{chiboth}
\end{eqnarray}

\vspace{-5mm}

In (\ref{chiboth}), we have considered that both background terms
vanish ($r_{0}=\widetilde{r}_{0}=0$), since this gives more stable
fits. Fitting the respective numerical data for both routes of
susceptibility to (\ref{chiboth}), the ratio $(\gamma / \nu)$ was
estimated and their values appear in
Tables~\ref{tablesample},\hspace{-2mm}
 ~\ref{tablesummation} for the same dilutions, see Fig.~\ref{Fig4}. In both
Tables, the ratio ($\gamma / \nu $) retains its pure Ising model
value ($\gamma / \nu = 7/4$) independently of dilution, thus
corroborating the weak universality hypothesis.

\begin{figure}[htbp]
\includegraphics*[height=0.3\textheight]{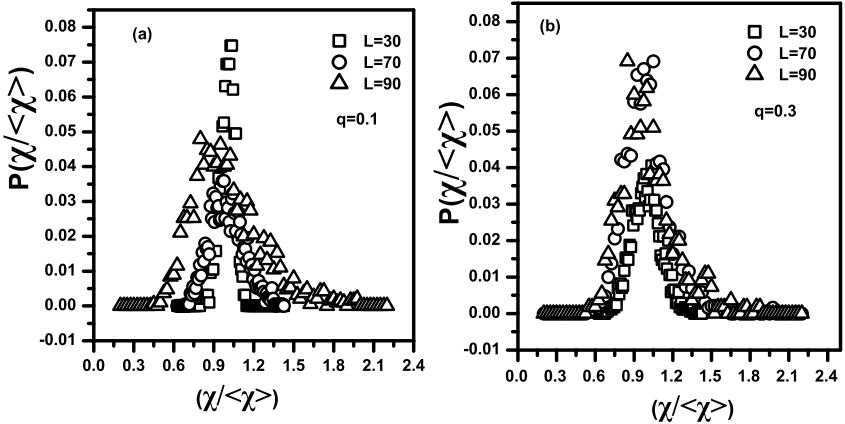}
\caption{\label{Fig5} Probability distribution of the relative
susceptibility  vs relative susceptibility $\chi _{r}$ ($\chi
_{r}$=$\chi _{j} /\! \left<\chi\right>$, $\chi _{j}$,
$\left<\chi\right>$ are the j-bin and global average
susceptibility, respectively) for dilutions $q=0.1(a), 0.3(b)$ and
lattice linear size L=30, 70, 90. The non-self-averaging behavior
is evident from the persistence of the width of the plots as the
lattice linear size $L$ increases.}
\end{figure}

An important issue arising in a disordered system is the notion of
self-averaging, that is, to what extent properties of the system
depend on the particular realization of the quenched random
variables, implying that the distribution of an observable becomes
sharper as $L$ increases. In case their relative widths remain
constant as $L \rightarrow\infty$, then we say the system is
non-self-averaging, \cite{wisemandomany}. In the pure Ising model
the respective distributions transform into delta functions as
$L\longrightarrow \infty$. To investigate the possibility of such
a behavior in the current disordered model, we produce $N_{s}$
samples of an observable $X$ (e.g., specific heat, susceptibility)
and form the respective probability density distribution
$P(X_{j}/\left<X\right>)$, where $X_{j}$ and $\left<X\right>$ are
the value of the observable in the $j$-bin and its global average,
respectively. If we identify $X$ with system's susceptibility
$\chi$, then the resulting probability density
$P(\chi_{i}/\!\left<\chi\right>)$ as a function of
$(\chi_{i}/\!\left<\chi\right>)$ for $q = 0.1, 0.3$ and $L = 30,
70, 90$ appears in Fig.~\ref{Fig5}. The respective curves of the
relative susceptibility for a specific value of dilution collapse
on each other irrespective of the value for $L$ and they do not
become sharper on increasing $L$. The general shape of the curves
remains the same independently of the value for $q$. This behavior
implies non-self-averaging. Similar behavior was also displayed by
the respective plots of pseudocritical temperatures and heat
capacity.

\vspace{-5mm}

\section{Conclusions and discussions}

\vspace{-4mm}

We have presented MC numerical data on the $2D$ RSDIM for three
cases of dilution spanning a wide range of dilution, $q = 0.1,
0.2, 0.3$, i.e., weak, intermediate and strong, for lattices with
linear size in the range $[20,120]$, using FSS and following the
Wang-Landau algorithm for the calculation of the density of
states. The calculations dealt with the estimation of the critical
temperature, the critical exponent $\nu$ and ratios ($\alpha /
\nu$) and ($\gamma / \nu$). The results indicate that as $q$
increases $\nu$ increases, while ($\gamma / \nu$) remains constant
and ($\alpha / \nu$) decreases. A consequence of the invariance of
($\gamma / \nu$) is that the exponent $\eta = 2 - (\gamma / \nu)$
is invariant, as well. Using the Rushbrooke equality $\alpha + 2
\beta + \gamma=2$ in conjunction with hyperscaling relation
$\alpha + d \nu = 2$, we deduced that ($\beta / \nu$) is also
invariant, retaining its pure Ising value, $\beta / \nu = 1/8$.
Referring to the critical temperature, although there exist four
different sequences, resulting from the various procedures for any
dilution $q$, each one exhibits a decreasing tendency, tending to
zero as $q$ tends to the percolation limit $q_{c} = 0.407255$.

A notable feature of the current-model data is the asymptotic
behavior of the specific heat as $L$ increases; it tends to a
saturation value for any $q$, in contradistinction to its
counterpart in the pure model that diverges. The other critical
quantity, susceptibility, still diverges as $L \rightarrow
\infty$.

An important issue is that if impurity concentrations larger than
$q=0.3$ towards the percolation limit $q_{c}$ are considered, then
the high dilution shall reduce significantly the long-range
correlations enhancing in this way the finite-size effects; to
moderate this effect larger lattices are needed at the cost of
larger computer-execution time.

Our results and findings are supporting the argument that the $2D$
RSDIM does not belong to the same universality class as the pure
$2D$ Ising model but to a new one. In conclusion, the present
results favor the weak universality scenario.

\vspace{-6mm}

\noindent
\ack{This research was supported by the Special Account
 for Research Grants of the University of Athens $\left(E\Lambda
 KE\right)$ under Grant No. 70/4/4096.}
\vspace{-5mm}



\end{document}